# An evolving network model with modular growth*

Zou Zhi-Yun(邹志云), Liu Peng(刘 鹏)[†], Lei Li(雷 立), and Gao Jian-Zhi(高健智)

*School of Civil Engineering & Mechanics, Huazhong University of Science and Technology,*

*Wuhan 430074, China*

In this paper, we propose an evolving network model growing fast in units of module, based on the analysis of the evolution characteristics in real complex networks. Each module is a small-world network containing several interconnected nodes, and the nodes between the modules are linked by preferential attachment on degree of nodes. We study the modularity measure of the proposed model, which can be adjusted by changing ratio of the number of inner-module edges and the number of inter-module edges. Based on the mean field theory, we develop an analytical function of the degree distribution, which is verified by a numerical example and indicates that the degree distribution shows characteristics of the small-world network and the scale-free network distinctly at different segments. The clustering coefficient and the average path length of the network are simulated numerically, indicating that the network shows the small-world property and is affected little by the randomness of the new module.

**PACC:** 0590, 0200

## 1 Introduction

Complex networks are currently being studied in many fields of science and engineering. Many systems in nature can be described by models of complex networks, such as interpersonal relationship network, scientific collaboration network, computer network, protein network, transportation network, electric power network, etc [1].

*Project supported by National Natural Science Foundation of China (Grant No. 51078165), Fundamental Research Funds for Central Universities (Grant No. HUST 2010MS030).
[†]Corresponding author. Email: lpengn@yahoo.cn

Complex network can describe the evolution mechanisms and statistical properties of systems based on graph theory and statistical physics. The empirical study on the real complex networks [2], indicated that the most networks showed small-world property and had a power-law degree distribution, while the small-world property means that the networks have a high degree of clustering coefficient and a small average path length. In 1998, Watts and Strogatz proposed WS small-world network model (WS model) [7], which realizes the transition from regular network to random network.. This model describes the small-world property successfully. However, the degree distribution of nodes for this model doesn't conform with the real networks. The BA scale-free network model (BA model) proposed by Barabási and Albert in 1999 [9], constructs a network with the power-law distribution of nodes degree. However, the clustering coefficient of the BA model is small for large-size network, indicating this model doesn't show significant small-world property.

In addition to the small-world property and the scale-free property, many real networks have module structure [10], which is consist of several modules. The connections are dense in one module, but are sparser between two modules. In the aspect of the research on network modeling with module structure, Li and Maini (2005) proposed an evolving model of scale-free network with community structure [19]. Nakazako et al (2007) proposed an evolving network model with the combination of preferential and non-preferential attachments [20]. Cui D et al (2009) studied the properties of asymmetrical evolved community networks [21]. Most evolving network models don't take the module structure into account. The research on network modeling with module structure is still lacking at present.

In the evolving rules of the BA model, network growth is usually achieved by adding only one node at each time step. But the network size may grow faster in real world [22][23], That is, several nodes may get connected each other to constitute a module, and join in the network at each time step. For instance, in scientific collaboration network, each group including several numbers can be regarded as a module and cooperates with other groups.

In this paper, we propose an evolving network model with modular growth, which shows the small-world property, the scale-free property and the module feature at the same time. At each time step several nodes added constitute a module (Hereinafter referred to as the new module), which shows the small-world property. A network grows fast in module units, between which the nodes are linked by preferential attachment on degree of nodes. Based on the mean field theory and a numerical example, we study the network statistic properties, including the degree distribution, the modularity measure, the clustering coefficient, and the average path length, and finally we discuss the small-world property and the scale-free property of the network.

## 2 Network model

### 2.1 Network morphology of the new module

In a real network, the module itself usually shows some network morphology. For example, in scientific collaboration network, the members may be fully connected within a team, which is a fully connected module, as shown in Fig. 1 (a); in urban road network, the roads are connected by many random factors within a new planned community, which shows a high randomness, as shown in Fig. 2 (b).

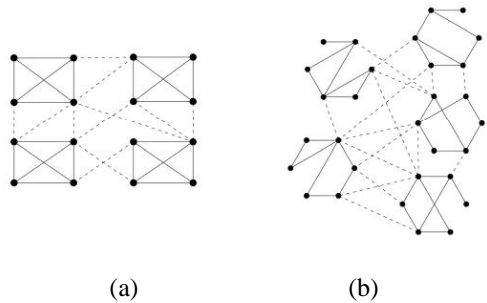

(a)  (b)

**Fig. 1.** Schematic diagram of the network morphology of the module

Under the influence of the randomness, most modules between full regular network and full random network [1], show small-world property significantly. Due to including less nodes, the new module generally doesn't show significant scale-free property. Therefore, we take WS small-world network as the network morphology of the new module in this paper. In the WS model, $a$ is the random rewiring probability, which can be adopted to adjust network randomness [7], as shown in Fig. 2. When

$a=0$, the new module shows regular network form; When $a=1$, the edges of the new module are all rewired randomly which shows ER random network form; When $0<a<1$, the new module is a small-world network, which lies between regular network and random network. With an increase in the random rewiring probability $a$, the randomness of the new module increases gradually.

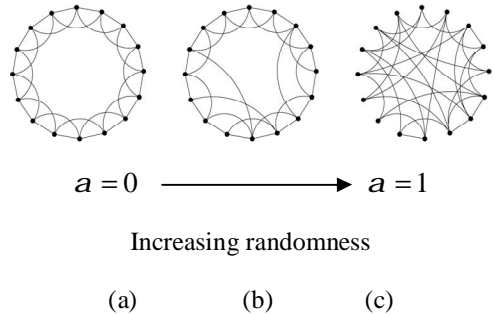

$a=0 \longrightarrow a=1$

Increasing randomness

(a)　　　(b)　　　(c)

**Fig. 2.** The random rewiring procedure of WS model

(a) regular network; (b) small-world network; (c) random network

## 2.2 Evolving progress

Unlike the previous evolving network models, the proposed model has a higher growth rate. At each time step, a module composed of several nodes is added to the network, and will experience a complete translation into a small world network (Both the regular network and random network can be regarded as the limited morphology of small-world network). After that, the nodes of the new module would connect to the nodes of the network by preferential attachment.

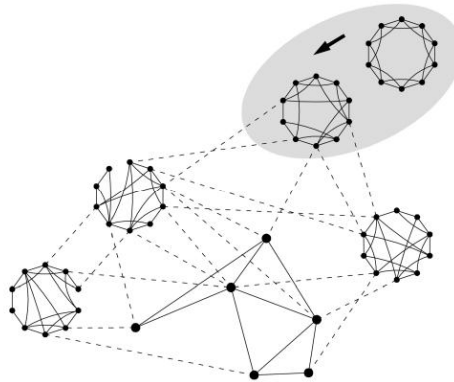

**Fig. 3.** The evolving progress of a network with modular growth

Fig. 3 shows the evolution progress of a network with modular growth. In the initial network, the number of nodes is $e_0=6$, and the number of edges is $m_0=9$. In each module, the number of nodes is $s = 10$, and the average degree of nodes is $K = 4$. The new module would get connected to some nodes of the network, and the number of connections is $m = 5$.

The nodes of the new module are chosen based on the mechanism of preferential attachment that the nodes with larger degree priority would get connected to the existing nodes. The existing nodes are also chosen by preferential attachment, and get connected to the new module.

The specific evolving process of the proposed network model is presented as follows:

(i) Constructing a new module with small world network morphology

(a) To start with order: Add $s$ nodes constituting a regular ring lattice, in which each node is connected to the $K$ closest nodes (there are $K/2$ nodes in each side), where $K$ is even. This lattice provides n edges where $n = Ks/2$.

b) To randomize: Randomly rewire each edge of the lattice with probability $a$ such that self-connections and duplicate edges are excluded. By changing $a$ one can closely monitor the transition between order $(a = 0)$ and randomness $(a = 1)$.

(ii) Network growth

Starting with a small number ($m_0$) of nodes and a small number ($e_0$) of edges, at every time step, add a new module of small-world network containing $s$ nodes and $n$ edges, which would get connected to $m$ existing nodes in the network. After $t$ time steps, in the network, the total number of nodes is $(m_0 + st)$ and the total number of edges is $[e_0 + (m + n)t]$.

(iii) Preferential mechanism of the new module attaching to the entire network

(a) To choose the nodes of the new module preferentially: When choosing the nodes of the new module which would connect to the nodes existed in the network, we assume that the probability $\prod(k_i)$ that node $i$ was chosen in the module depends on its degree $k_i$:

$$\prod(k_i) = \frac{k_i}{\sum_x k_x} \tag{1}$$

(b) To choose the nodes existed in the network preferentially: When choosing the nodes existed in the network which would connect to the chosen nodes of the new modules, we assume that the probability $\prod(k_j)$ that node $j$ was chosen in the network depends on its degree $k_j$:

$$\prod(k_j) = \frac{k_j}{\sum_y^{m_0+st} k_y} \tag{2}$$

## 3 Network properties

### 3.1 Modularity measure

There is modular structure in the network model proposed, modularity measure is proposed to describe the strength of module structure in a network. According to the Newman and Girvan's study [11]:

$$Q = \sum_r (e_{rr} - a_r^2) \tag{3}$$

where $e_{rr}$ indicates the fraction of edges that links the vertices in module $r$ to the vertices in module $r$, $a_r$ indicates the fraction of edges that links to the vertices in module $r$. For the network model proposed, we can write

$$e_{rr} = \frac{n}{m_0 + (n+m)t} \tag{4}$$

$$a_r = \sum_j e_{rj} = \frac{n+m}{m_0 + (n+m)t} \tag{5}$$

then

$$Q = \frac{n}{m_0/t + n + m} - \frac{n^2}{(m_0/t)^2 + (n+m)t} \approx \frac{n}{n+m} \tag{6}$$

When the time step $t$ is large enough so that the network size is large, $m_0$ and the second formula can be ignored. Then we can adjust modularity measure by changing ratio of inner-module edges $n$ and inter-module edges $m$.

Take the initial network as shown in Fig. 3 as an example, the related parameter can be given by: $e_0 = 6$, $m_0 = 9$, $s = 10$, $K = 4$ and $n = 20$. Keeping $n = 20$ unchanged, the network modular measure $Q$ is respectively equal to 0.8, $\frac{2}{3}$, 0.5, and $\frac{1}{3}$ as $m = 5, 10, 20, 40$, indicating that the modular measure $Q$ decreases with $m$ increasing.

### 3.2 Degree distribution

In this paper, we analyse the degree distribution of the new model using the mean-field theory, by assuming that the change of node degree is continuous. When the time step is large enough, the initial network size is negligible, and the change rate of node degree can be expressed as

$$\frac{\partial k_i}{\partial t} = m \frac{k_i}{\sum k_l} = m \frac{k_i}{2m_0 + 2(n+m)t} \approx \frac{m}{2(n+m)} \cdot \frac{k_i}{t} \qquad (7)$$

Assuming the initial value $k_i(t_i) = k_x$, $k_x$ represents the initial degree of node $i$ belonged to the new module after being added to the network, Equation(7) leads to

$$k_i(t) = k_x \left(\frac{t}{t_i}\right)^{\frac{m}{2(n+m)}} \qquad (8)$$

The probability density function of time $t_i$ can be written as $P_i(t_i) = 1/(t + \frac{e_0}{s})$, therefore we can obtain the probability that the degree of node $i$ of the network is equal to $k$:

$$P(k) = \frac{\partial P(k_i(t) < k)}{\partial(k)} \approx \left(\frac{2n}{m} + 2\right) \cdot (k_x)^{\frac{2n}{m}+2} \cdot k^{-(\frac{2n}{m}+3)} \qquad (9)$$

where $k_x \leq k$.

The probability that the initial degree of node $i$ is $k_x$ depends on the small-world evolution and preferential mechanism in the new module. According to the Newman and Watts's study [8] on the WS model, when $a = 0$, all the edges of the new module wouldn't be rewired, and the degree distribution of the new module is the same as regular network, which are all $K$; when $a \geq 0$, each edge keeps an endpoint unchanged in the process of rewiring, so each node is linked by $K/2$ edges at least. Therefore after the small-world evolution of the new module, the probability that the degree of node $i$ of the new module is $k'$ can be written as

$$u(k') = \begin{cases} 0 & , k' < K/2 \\ \sum_{n=0}^{\min(k'-K/2, K/2)} C_{K/2}^n (1-a)^n a^{K/2-n} \frac{(aK/2)^{k'-K/2-n}}{(k'-K/2-n)!} e^{-aK/2} & , k' \geq K/2 \end{cases} \quad (10)$$

And then the nodes of the new module will be chosen preferentially, that would be repeated $m$ times. The probability that a node with $k'$ degree is chosen $k''$ times (namely its degree increases by $k''$, and $k'' \leq m$) can be written as

$$w(k'') = C_m^{k''} (\frac{k'}{sK})^{k''} (\frac{sK-k'}{sK})^{m-k''} \quad (11)$$

Then, we can obtain the probability that the initial degree of the node $i$ of the new module is $k_x$ after the node $i$ is added into the network:

$$p(k_x) = \begin{cases} C_m^{k_x-K} (\frac{1}{s})^{k_x-K} (\frac{s-1}{s})^{m-(k_x-K)} & , a = 0 \\ \sum_{k'=K/2} [u(k') \cdot w(k_x - k')] & , 0 < a \leq 1 \end{cases} \quad (12)$$

As seen from Equation (12), the initial degree of node $i$ of the new module is $k_x$ after the node $i$ is added into the network. The probability equation is similar to Binomial distribution, related to the small-world evolution and the preferential mechanism in the new module.

According to Eqs. (5) and (8), the probability that the degree of node $i$ in the network is $k$ can be written as

$$P(k) = A \cdot k^{-(A+1)} \cdot f(k) - k^{-A} \cdot \frac{\partial f(k)}{\partial k} \quad (13)$$

where $A = \frac{2n}{m} + 2$, and $f(k) = \sum_{k_x=K/2}^{k}(k_x)^A p(k_x)$. The formula $f(k)$ contains factorial expression, which is very difficult to derivative, so we keep the expression $\frac{\partial f(k)}{\partial k}$ in formula $P(k)$.

Fig. 4 shows the analytical results of $f(k)$ with double logarithmic coordinates, $f(k)$ increases with $k$ increasing, but the increasing rate gradually decreases. Under the assumption of the continuous change of node degree $k$, when the node degree $k$ is small, there is $\frac{\partial f(k)}{\partial k} > 0$, and the probability $P(k)$ doesn't obey the power-law distribution. When the node degree $k$ is large, $f(k)$ remains unchanged, there are $f(k) \approx \sum_{k_x=K/2}(k_x)^A p(k_x)$, $\frac{\partial f(k)}{\partial k} \approx 0$ and the probability $P(k) \approx [A \cdot \sum_{k_x=K/2} k_x^A p(k_x)] \cdot k^{-(A+1)}$, which obeys a power-law distribution with exponent $g = A + 1 = \frac{2n}{m} + 3$, where $g > 3$ and depends on $m$, $n$.

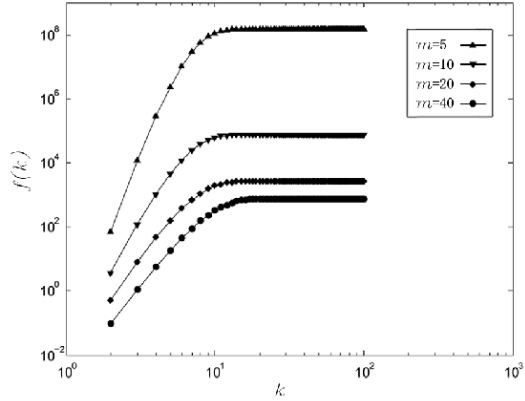

**Fig. 4.** Dependence of $f(k)$ on node degree $k$ with $m = 5, 10, 20, 40$.

In Fig. 5, comparative analysis is made with the analytical results of degree distribution and its numerical results in double logarithmic coordinates. For simplicity, let $\frac{\partial f(k)}{\partial k} = 0$ and $P(k) = A \cdot k^{-(A+1)} \cdot \sum_{k_x=K/2}^{k}(k_x)^A p(k_x)$. Under this condition, when the node degree $k$ is small, the analytical results showed in figure are too large compared with the accurate analytical results. However, when the node degree $k$ is large, the analytical results showed in figure are close to the accurate analytical results. As shown in Fig. 5, we find that when the node degree $k$ is small, the analytical results are larger than the numerical results significantly, and with the node degree $k$ increasing, the analytical results are close to the numerical results gradually. Therefore, we assert that the accurate analytical results are consistent to the numerical results.

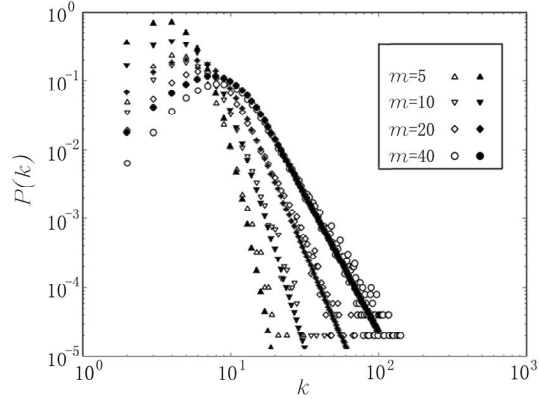

**Fig. 5.** Degree distribution with $m = 5, 10, 20, 40$ and the related parameters $N = 5000$, $a = 0.5$, $e_0 = 6$, $m_0 = 9$, $s = 10$, $K = 4$, and $n = 20$. The filled marks represent the analytical results, while the unfilled marks represent the numerical results.

Fig. 5 indicates that the degree distribution shows characteristics of the small-world network and the scale-free network distinctly at different segments. When the node degree $k$ is small, the probability $P(k)$ is similar to the degree distribution of the small world network model, and there is a significant peak at the node degree $k = [K + \frac{m}{s}]$, and then the probability $P(k)$ gradually decreases with the increase in node degree. However, when the node degree $k$ is large, the probability $P(k)$ obeys the power-law distribution, and the exponent $g$ decreases from 11 ,7 ,5, to 4,with the increase in $m$ from 5, 10, 20 to 40.

### 3.3 Clustering coefficient and average path length

Clustering coefficient is a statistic measuring network clustering, which is defined as the ratio of the actual number of edges among the given neighbor nodes, to theoretical number of edges [1]. For a selected node $i$ in the network, there are $k_i$ edges linked to other nodes $k_i$. Among the $k_i$ edges there are actually existing $E_i$ edges. The clustering coefficient of node $i$ is then

$$C_i = \frac{2E_i}{k_i(k_i-1)} \tag{14}$$

The clustering coefficient $C$ of the network with $N$ nodes is the average of $C_i$:

$$C_i = \frac{1}{N}\sum_{i=1}^{N} C_i \tag{15}$$

Average path length is an important statistic characteristic. Small average path length means that the network shows small-world property [1]. For a connected undirected network, path length from node $i$ to node $j$ is defined as the number of edges on the shortest path between them, while the average path length $L$ is defined as the average value of path lengths between any two nodes in a network:

$$L = \frac{1}{\frac{1}{2}N(N+1)}\sum_{i\geq j} d_{ij} \tag{16}$$

We study the dependence of clustering coefficient and average path length on the probability $a$ (from 0 to 1) and the network size $N$ (from 500 to 5000) with the increase in m from 5, 10, 20, to 40. The related parameters of the network are: $e_0 = 6$, $m_0 = 9$, $s = 10$, $K = 4$, and $n = 20$. The numerical results for a series of computer simulations using the model are shown in Figs. 6 and 7. In Fig. 7 logarithmic abscissa is used.

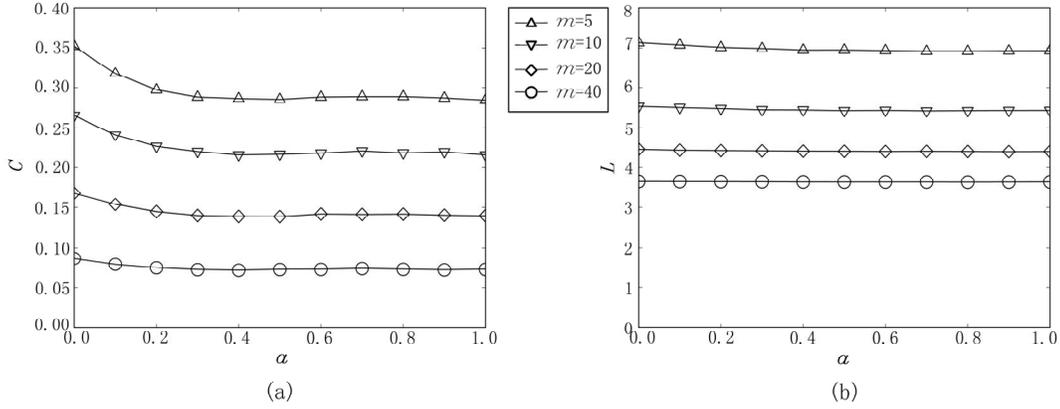

**Fig. 6.** Figure 6(a): Dependence of clustering coefficient on the probability $a$ with $N = 5000$, and $m = 5, 10, 20, 40$, respectively. Figure 6(b): Dependence of average path length on the probability $a$ with $N = 5000$, and $m = 5, 10, 20, 40$, respectively.

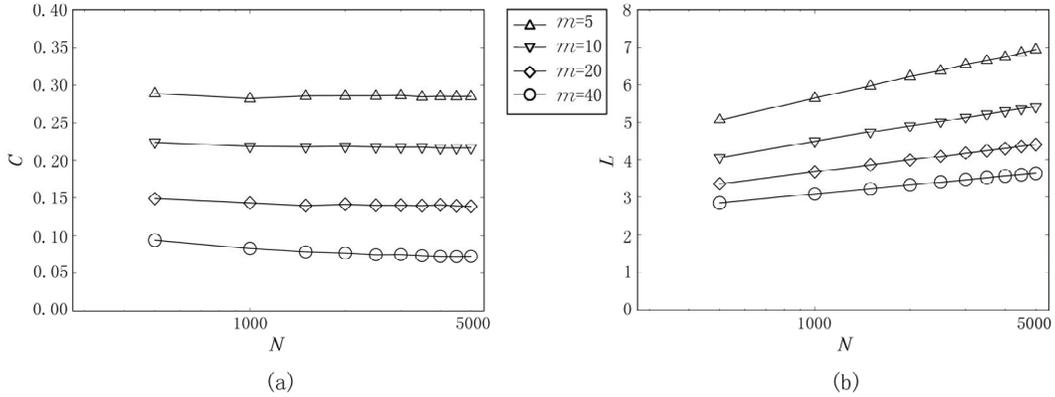

**Fig. 7.** Figure 7(a): Dependence of clustering coefficient on network size $N$ with $a = 0.5$, and $m = 5, 10, 20, 40$, respectively. Figure 7(b): Dependence of average path length on network size $N$ with $a = 0.5$, and $m = 5, 10, 20, 40$, respectively.

As shown in Fig. 6, the probability $a$ for this case keeps at 0.5. With the network size increasing, clustering coefficient remains at a high level, and average path length grows in logarithmic manner. This indicates that the network proposed shows small-world property.

As shown in Fig. 7, the network size for this case keeps at 5000. With probability $a$ increasing, average path length remains essentially unchanged. When the probability $a$ ranges at $[0,0.3]$, clustering coefficient decreases slightly with $a$ increasing, but remains at a high lever when the probability $a$ increases to 0.3. Overall, the value of the probability $a$ doesn't significantly alter clustering coefficient and average path length. That is, the small world property of the network has little effect on the randomness of the new module.

## 4 Conclusions

Module structure exists in many real networks, which grow not always in units of node but sometimes in units of module. The module usually shows some network morphology.

In this paper we propose an evolving network model with modular growth. Unlike most of the models studied in the past, the new model grows faster, and several interconnected nodes constituting a module of small-world network, join in the network at each time step. We study the modularity measure of the proposed model, which can be adjusted by changing ratio of inner-module edges $n$ and inter-module edges $m$. Based on the mean field theory, we develop an analytical function of the degree distribution, which is verified by a numerical example and indicates that the degree distribution shows characteristics of the small-world network and the scale-free network distinctly at different segments. The degree distribution is similar to the small-world model for small node degree $k$, and is a power-law distribution for large node degree $k$. The clustering coefficient and the average path length are simulated numerically. The results indicates that the network shows the small-world property, and is affected little by the randomness of the new module.